\documentclass[floatfix,aip,reprint]{revtex4-1}
\usepackage[utf8]{inputenc}
\usepackage[T1]{fontenc}
\usepackage{textcomp}
\usepackage[euler]{textgreek}
\usepackage{gensymb}
\usepackage[caption=false]{subfig}
\usepackage[section]{placeins}
\usepackage{graphicx}
\usepackage{siunitx}
\usepackage{paralist}
\usepackage{booktabs}
\usepackage{miller}
\usepackage[version=3]{mhchem}
\usepackage{color}
\usepackage{pst-all}

\begin{document}

\hyphenation{pseu-do-mor-phic InGaN GaN ca-tho-do-lu-mi-nes-cen-ce se-mi-bulk tri-me-thyl-in-di-um tri-me-thyl-gal-li-um}

\title{Nano selective area growth of GaN by MOVPE on 4H-SiC using epitaxial graphene as a mask: towards integrated III-nitride / graphene / SiC electronics and optoelectronics.}

\author{Renaud Puybaret}\email{renaudpuybaret@gmail.com}
\affiliation{School of Electrical and Computer Engineering - Georgia Institute of Technology - 30332 Atlanta GA - USA}
\affiliation{Georgia Institute of Technology - CNRS UMI 2958 -  2 rue Marconi 57070 Metz - France}

\author{Gilles Patriarche}
\affiliation{CNRS, Laboratoire de Photonique et de Nanostructures - Route de Nozay 91460 Marcoussis -  France}

\author{Matthew B. Jordan}
\affiliation{School of Electrical and Computer Engineering - Georgia Institute of Technology - 30332 Atlanta GA - USA}
\affiliation{Georgia Institute of Technology - CNRS UMI 2958 -  2 rue Marconi 57070 Metz - France}

\author{Suresh Sundaram}
\affiliation{Georgia Institute of Technology - CNRS UMI 2958 -  2 rue Marconi 57070 Metz - France}

\author{Youssef El Gmili}
\affiliation{Georgia Institute of Technology - CNRS UMI 2958 -  2 rue Marconi 57070 Metz - France}

\author{Jean-Paul Salvestrini}
\affiliation{Universit\'e de Lorraine, CentraleSup\'elec, LMOPS - EA4423 57070 Metz - France}

\author{Paul L. Voss}
\affiliation{School of Electrical and Computer Engineering - Georgia Institute of Technology - 30332 Atlanta GA - USA}
\affiliation{Georgia Institute of Technology - CNRS UMI 2958 -  2 rue Marconi 57070 Metz - France}

\author{Walt A. de Heer}
\affiliation{School of Physics - Georgia Institute of Technology - 30332 Atlanta GA, USA}  

\author{Claire Berger}\email{claire.berger@physics.gatech.edu}
\affiliation{School of Physics - Georgia Institute of Technology - 30332 Atlanta GA, USA}  
\affiliation{CNRS, Institut N\'eel - BP166 38042 Grenoble Cedex 9 - France}

\author{Abdallah Ougazzaden}\email{aougazza@georgiatech-metz.fr}
\affiliation{School of Electrical and Computer Engineering - Georgia Institute of Technology - 30332 Atlanta GA - USA}
\affiliation{Georgia Institute of Technology - CNRS UMI 2958 -  2 rue Marconi 57070 Metz - France}

\date{\today}

\begin{abstract}

{\textit{Abstract}: We report the growth of high-quality triangular GaN nanomesas, 30-nm thick, on the C-face of 4H-SiC using nano selective area growth (NSAG) with patterned epitaxial graphene grown on SiC as an embedded mask. NSAG alleviates the problems of defective crystals in the heteroepitaxial growth of nitrides, and the high mobility graphene film can readily provide the back low-dissipative electrode in GaN-based optoelectronic devices. The process consists in first growing a 5-8 graphene layers film on the C-face of 4H-SiC by confinement-controlled sublimation of silicon carbide. The graphene film is then patterned and arrays of 75-nanometer-wide openings are etched in graphene revealing the SiC substrate. 30-nanometer-thick GaN is subsequently grown by metal organic vapor phase epitaxy. GaN nanomesas grow epitaxially with perfect selectivity on SiC, in openings patterned through graphene, with no nucleation on graphene. The up-or-down orientation of the mesas on SiC, their triangular faceting, and cross-sectional scanning transmission electron microscopy show that they are biphasic. The core is a zinc blende monocrystal surrounded with single-crystal hexagonal wurtzite. The GaN crystalline nanomesas have no threading dislocations, and do not show any V-pit. This NSAG process potentially leads to integration of high-quality III-nitrides on the wafer scalable epitaxial graphene / silicon carbide platform.}

\end{abstract}

\maketitle

For the past five years there has been much interest in integrating graphene and nitride-based semiconductors. For instance boron nitride is used as a substrate and top gate for graphene electronics \cite{giovanetti2007,dean2010,geim2011,giovanetti2007}. Aluminium nitride and silicon nitride can both serve as a mask for selective epitaxial graphene growth on silicon carbide\cite{zaman2010,puybaret2015}. Graphene is also included in InGaN / GaN optoelectronic devices as a transparent top window electrode  \cite{jo2010,jeon2011,lee2011}, or as a substrate for growth of transferable GaN-InGaN-based multi-quantum well (MQWs) light-emitting diodes (LEDs) \cite{chung2010,kim2014}, and even as a heat-spreading underlayer\cite{han2013}.  

During bulk heteroepitaxy GaN and InGaN grow with many defects, notably threading dislocations leading to V-pits, lattice mismatch strain relaxation and transition from 2D to 3D growth and inhomogeneous indium incorporation (for InGaN), \cite{goh2009,puybaret2015_4}. Such defects are detrimental to optoelectronic (and electronic) device performance, as they act as non-radiative recombination centers and leakage-current paths \cite{speck1996,wu1998,nakamura2000}.
Heteroepitaxial nano selective area growth \cite{zubia1999,sun2004,goh2009} (NSAG) was shown to alleviate these problems and produce high-quality and thick InGaN and GaN crystals. InGaN and GaN nanopyramids, with thickness over 100 nm and with no defects have been consistently grown with high throughput on various substrates, including silicon \cite{puybaret2014,puybaret2015_2,puybaret2015_4}. Difficulty with NSAG come from the additional processing steps required to pattern the mask and remove it after the nitride growth, and to implement the device back electrodes \cite{jeon2011,kang2013,song2014,chang2015}. 

In this letter, we report on a new NSAG technique that uses multilayer epitaxial graphene (MEG) on SiC as an embedded mask, which could then be used the back electrode to the grown GaN. This is possible because, as we show below, GaN solely grows on SiC from the holes etched through the graphene and as it becomes thicker than the depth of the few-atomic-layer hole, it makes a recovering contact with graphene.

This process is a simple and straightforward approach for combining III-nitride crystals with epitaxial graphene on silicon carbide. Epitaxial graphene on SiC is an ideal platform for electronics and optoelectronics fabrication. It can be grown in large continuous films on silicon carbide \cite{deheer2011_3,palmer2014}, making it a wafer-scalable technology: indeed SiC wafers are available commercially up to 150 mm (6 inches) in diameter, and price is dropping steadily down to below 100 USD per square centimeter. Moreover epitaxial graphene on SiC is a high mobility, high quality, and scalable form of graphene \cite{berger2004,berger2006,sprinkle2010,novoselov2012,deheer2014}. The contact between GaN over epitaxial graphene enables the use of the honeycomb carbon material as a high electron mobility and high thermal conductivity back-electrode. The high electron mobility of graphene leads to higher thermal conductivity\cite{berger2006}, hence tackling a capital problem of large-scale integration of electronics and optoelectronics: the dissipation of heat \cite{intel2004}. This makes SiC is a substrate of choice for III-nitride growth, as exemplified by the mass production of white-light LEDs using InGaN-on-SiC technology \cite{doverspike2002patent}. 

Graphitized SiC substrates are therefore an interesting option for GaN technology. The process presented here circumvents the issue of lattice mismatch during heteroepitaxy by growing gallium nitride selectively on silicon carbide using nanoscale-patterned epitaxial graphene as a mask.  Wurtzite GaN has a small in-plane lattice mismatch (3.5\%) with hexagonal SiC and grows selectively on SiO$_2$-masked Si-face 6H-SiC \cite{goh2009}. Then SiC and epitaxial graphene are compatible with high temperature MOVPE growth of high quality GaN materials. Moreover, a selective area growth (SAG) of GaN on epitaxial-graphene-masked SiC is a compelling device design to provide a direct electronic connection between an active optoelectronic alloy and a high electron mobility \cite{berger2004,berger2006,deheer2014}, and high thermal conductivity \cite{balandin2008,han2013} electrode, potentially improving the performance of GaN-based nano-optoelectronic devices. Finally, the dual purpose of epitaxial graphene, both as mask for NSAG and potential back-electrode, drastically reduces the number of steps in the process, hence lowering both the cost and the environmental impact of the fabrication of nitride-based electronics and optoelectronics.

\section*{Results and Discussion}

\begin{figure}[h!]
\centering
\includegraphics[width=.8\columnwidth]{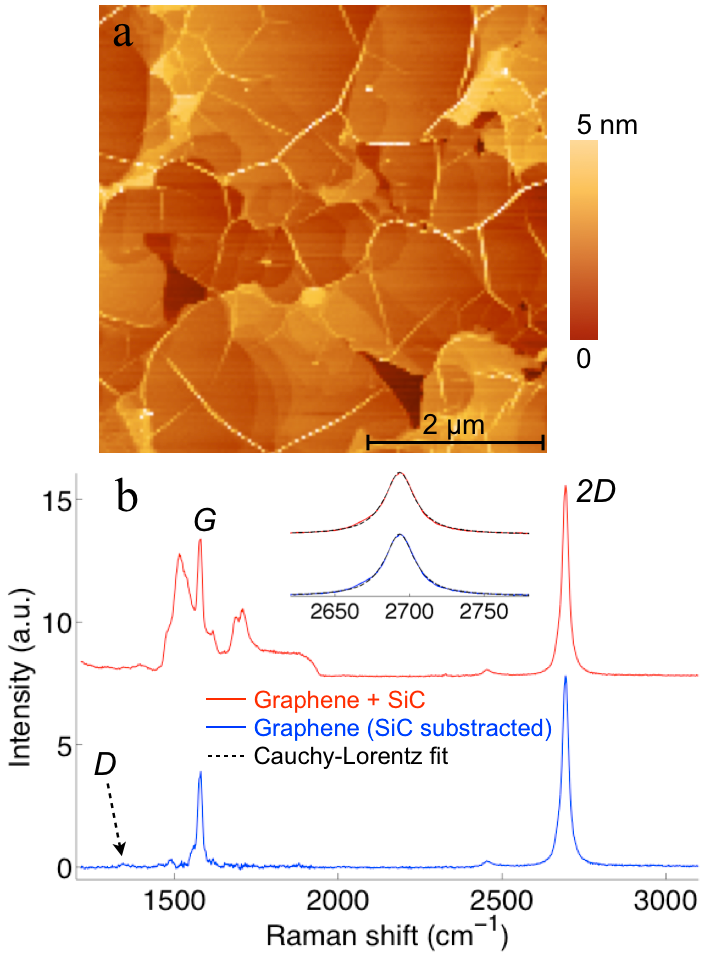}
\caption{After CCS graphitization of C-face SiC. (a) AFM image, 5~$\times$~5~\textmu m$^2$. (b) Raman spectra with and without the SiC background, respectively red and blue; the inset zooms the 2D peak.}
\label{fig:FIG1_ccs_cface_46SAG07}
\end{figure}

The first step in our process is the growth of graphene on the C-face of 4H-SiC by confinement-controlled sublimation (CCS) of silicon carbide \cite{deheer2011_3}. An atomic force microscopy (AFM) image of the surface of our graphene sample shown in Fig.\ref{fig:FIG1_ccs_cface_46SAG07}a confirms the presence of graphene layers, as shown by the MEG characteristic pleats. Typical characteristics in terms of pleat structure, including pleat height (1.5-2.4 nm), pleat surface density, and semi-hexagonal orientation, can be observed  \cite{deheer2011_3,hu2012}. The underlying SiC step structure is clearly visible, with continuous graphene layers draped over it.

The Raman spectrum (wavelength 532 nm) in Fig.\ref{fig:FIG1_ccs_cface_46SAG07}b reveals the characteristic graphene peaks. The graphene 2D and G peaks are clearly identified. The raw data (red) include the SiC contribution, that is substracted to give the signal of graphene only (blue). The 2D peak of all 5 samples measured so far can be fitted by a single Cauchy-Lorentz distribution \cite{faugeras2008} centered between 2694 and 2700~cm$^{-1}$ and with FWHM comprised between 23 and 41~cm$^{-1}$. In this study graphene thickness ranges from 5 to 8 layers, at 0.34 nm per layer (see STEM images in Figs.\ref{fig:FIG3_tem_all}-\ref{fig:FIG5_tem_cubic}).
The D peak at 1350 cm$^{-1}$ is very small, or unnoticeable, in all cases. This indicates low defect density in the graphene lattice. Specifically, we do not observe the characteristic shouldered 2D peak of highly ordered pyrolitic graphite (HOPG), as already reported for multilayered epitaxial graphene on the C-face \cite{faugeras2008}. 

The next step in the process is to etch patterns in the graphene using electron-beam lithography (with hydrogen silsesquioxane, HSQ, as the negative resist) and oxygen plasma reactive ion etching (RIE). HSQ is then removed in buffered oxide etch (BOE). BOE, also known as buffered hydrofluoric acid, serves a dual purpose here as it also removes the possible oxide of SiC formed during RIE etching. The substrate for GaN growth therefore consists of an array of 75-nanometer-wide openings in a continuous multilayer graphene film. Oxygen RIE, coupled with HF treatment, exposes SiC in the holes. The last step of the process is to grow 30-nm-thick GaN nanostructures by MOVPE, as described in Refs.\cite{martin2008}.

\begin{figure}[h]
\centering
\includegraphics[width=\columnwidth]{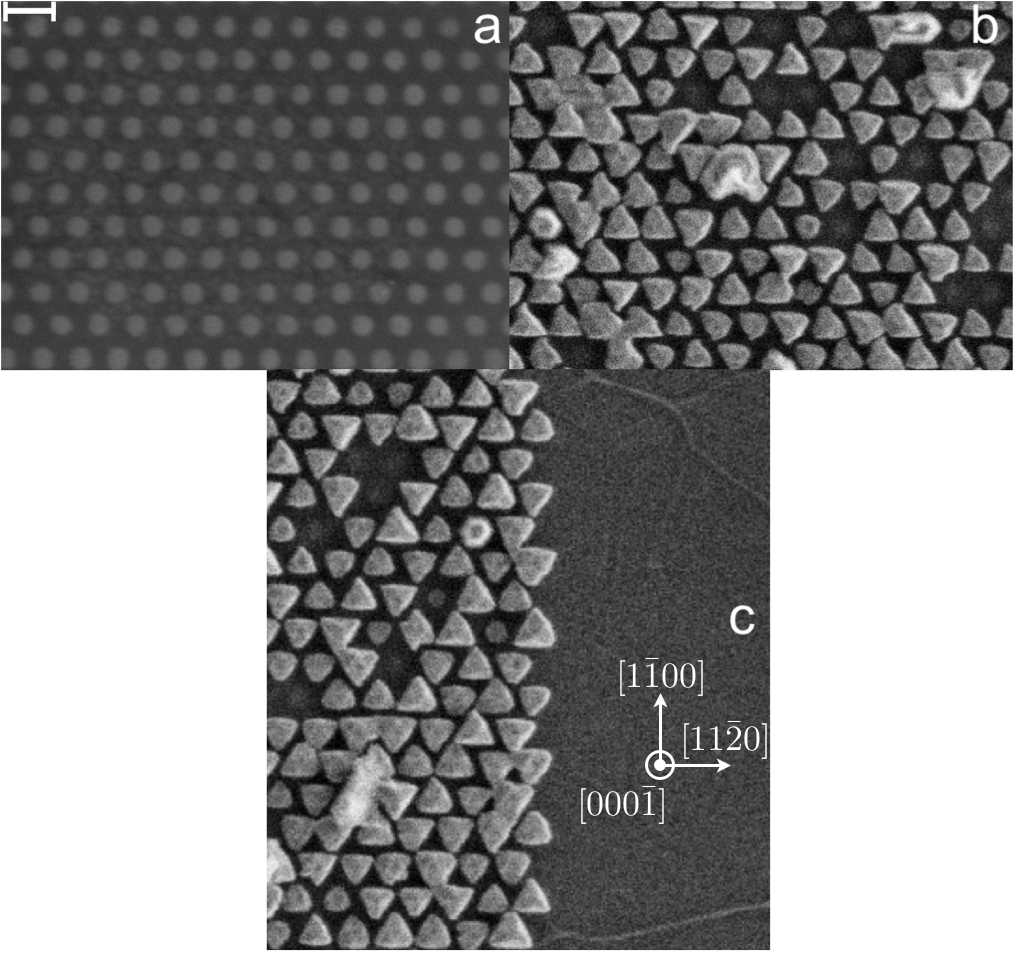}
\caption{Scanning electron microscope (SEM) images of NSAG of 30-nm-thick GaN on 4H-SiC C-face using epitaxial graphene as a mask. \textit{Scale bar in (a) is 200 nm, same scale in all panels.} (a) before GaN growth, pattern of graphene (dark) on SiC (clear). (b) after MOVPE of 30 nm of GaN, GaN triangular nanomesas (clear) grow selectively in the openings revealing the SiC, and not on the graphene (dark); all the nanomesas exclusively point up or down. (c) Zoom on the edge of the mask, where we can see that GaN nucleates only on the SiC, and does not nucleate at all on the graphene, not even on the pleats: the growth is perfectly selective; \textit{the vector basis in (c) describes the crystallographic orientations of the 4H-SiC substrate, same basis for all panels.}}
\label{fig:FIG2_graphene_nanosag_sem}
\end{figure}

\begin{figure}[h!]
\centering
\includegraphics[width=\columnwidth]{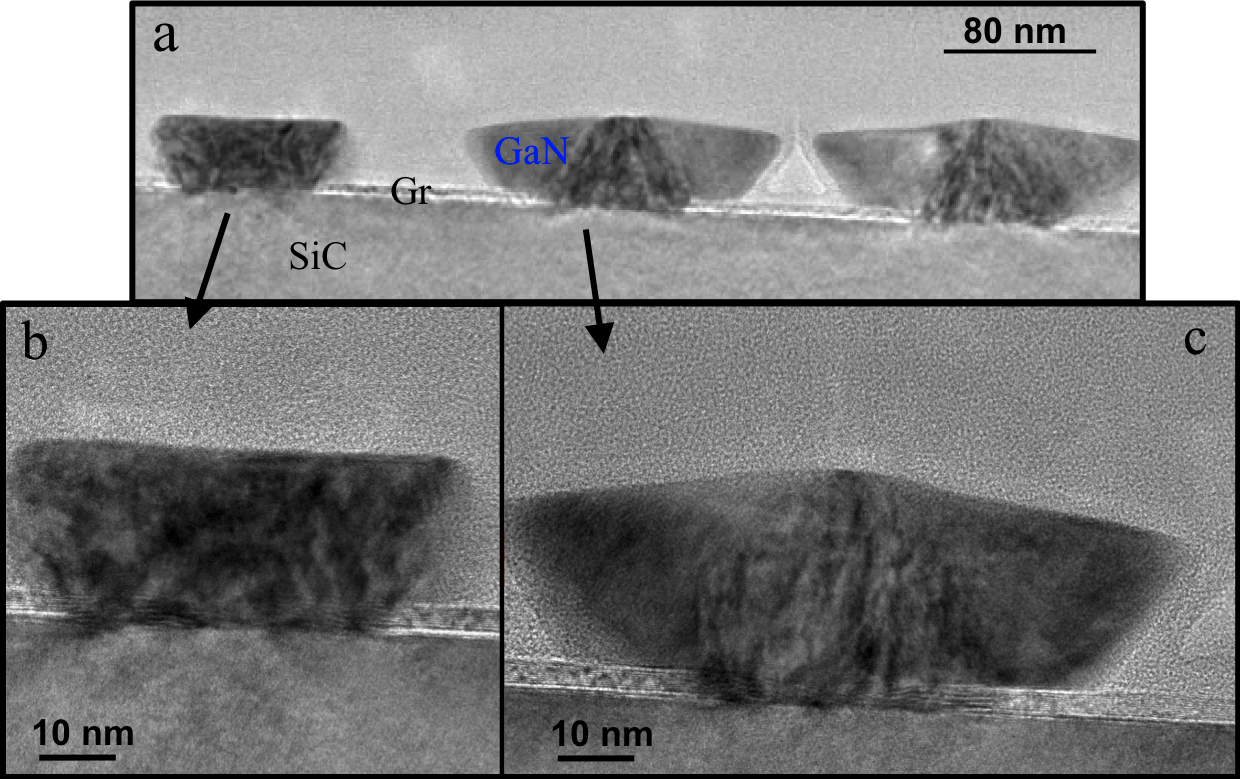}
\caption{Cross-sectional scanning transmission electron microscope (STEM) images: (a) 3 consecutive GaN nanomesas; (b) Zoom on the smaller left-hand nanomesa, with the flat top, having a mostly wurtzite lattice; (c) Zoom on the middle GaN nanomesa, having a large central cubic (a.k.a. zinc blende) seed on which hexagonal wurtzite facets have grown. \\ \textit{[Note] A mix of surface defects and Moir\'e interference patterns (c-GaN and/or w-GaN and/or graphene) can seen in (b) and (c): these imagery artifacts are due to the focused ion beam (FIB) cut necessary for STEM preparation; in particular the graphene is completely etched in the hole: the FIB sliver is wider than the hole in the graphene (70 nm), hence the graphene behind the hole is imaged by the STEM and overlaps with the c-GaN and/or w-GaN, forming Moir\'e interference fringes \cite{lechner2011,boda2015}.}}
\label{fig:FIG3_tem_all}
\end{figure}

\begin{figure}[h!]
\centering
\includegraphics[width=.9\columnwidth]{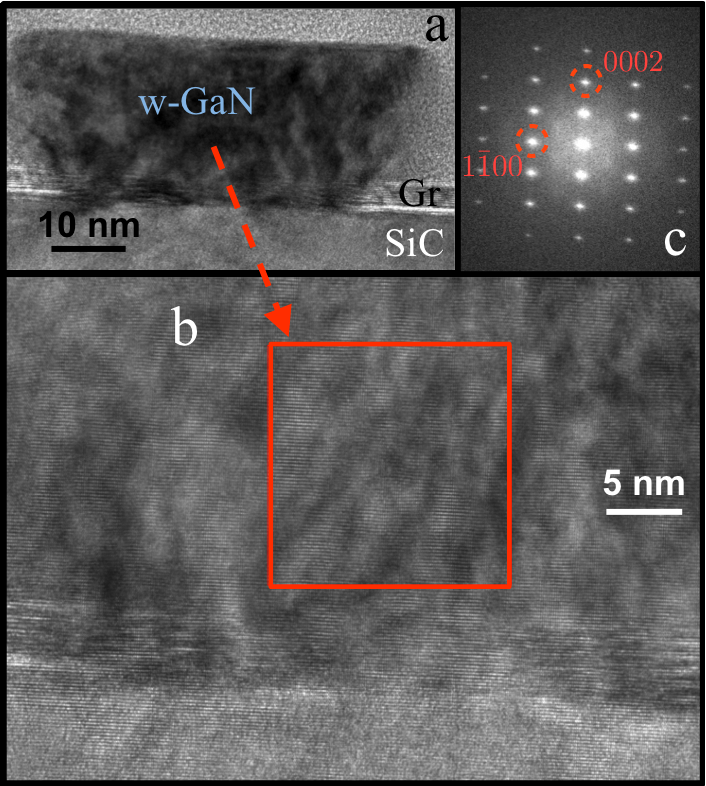}
\caption{Single-crystal wurtzite GaN nanomesa: (a) STEM image. (b) Zoom and definition of the zone (red square) taken for fast Fourier transform (FFT); \textit{see [Note] on Fig.\ref{fig:FIG3_tem_all}}. (c) FFT diffraction pattern of the area in the red box showing wurtzite GaN; \textit{local intensity maxima 0002 and 1\=100 make an angle of \textbf{90$^\circ$} relative to the origin, characteristic of the wurtzite lattice.}}
\label{fig:FIG4_tem_wurtzite}
\end{figure}

\begin{figure*}[h!]
\centering
\includegraphics[width=1.7\columnwidth]{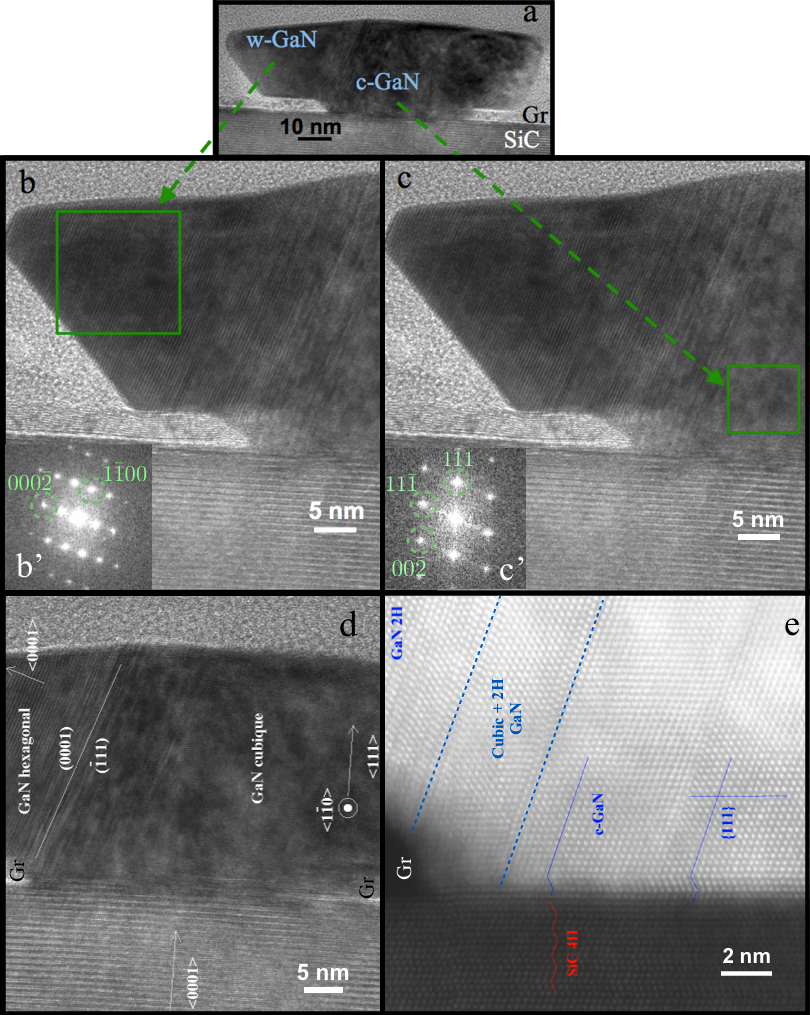}
\caption{GaN nanomesa with central cubic seed, and hexagonal facets (\textit{see [Note] on Fig.\ref{fig:FIG3_tem_all}}): (a) Cross-sectional STEM image; (b) Zoom and definition of the wurtzite zone (green box) taken for FFT and (b') FFT diffraction pattern of the area in the green box showing wurtzite GaN. (c) Zoom and definition of the cubic zone (green box) taken for fast Fourier transform (FFT) and (c') FFT diffraction pattern of the area in the green box showing cubic GaN, where 1\=10 faces the reader; \textit{local intensity maxima 111 and 11\=1 make an angle of \textbf{70.5$^\circ$} relative to the origin, characteristic of the zinc-blende / cubic lattice.} (d) Full STEM analysis and crystallographic directions. (e) Zoom on the SiC / GaN interface in the vicinity of the epitaxial graphene mask, high-angle annular dark field (HAADF) detector, showing both cubic and wurtzite phases on GaN, as well as the biphasic transition region; \textit{the visible atoms are Ga in GaN and Si in SiC.}}
\label{fig:FIG5_tem_cubic}
\end{figure*}

We observe that the GaN grows only from the SiC surface in the holes, cf Fig.\ref{fig:FIG2_graphene_nanosag_sem}, in the shape of equilateral-triangle nanomesas with no nucleation on the epitaxial graphene. We find that the GaN nanostructures on SiC are significantly different from the typical hexagonal nanopyramids and their six (1\=101) or (1\=102) r-plane facets \cite{goh2009,puybaret2014,puybaret2015_2,puybaret2015_4}. Here we use to our advantage the nonwetting property of graphene. The fact that coatings don't stick to graphene is in general problematic for device fabrication and nanoelectronics processing \cite{guo2013,dong2014}. The absence of GaN growth on MEG, not even on graphene pleats, underlines the high structural quality of the pristine MEG in this study.

A cross-section scanning transmission electron microscope (STEM) image of GaN nanomesas (see Fig.\ref{fig:FIG3_tem_all}) clearly shows that the growth of GaN does not emerge from the epitaxial graphene patterns. The single orientation and faceting of the GaN nanomesas in Figs.\ref{fig:FIG2_graphene_nanosag_sem} and \ref{fig:FIG3_tem_all} confirms epitaxial growth of GaN on SiC. 
The vast majority (four fifths) of GaN nanomesas have a large central triangular cubic seed, above which hexagonal wurtzite facets have grown. These mesas are the larger ones with a roof-shaped top as seen on Figs.\ref{fig:FIG3_tem_all}c and \ref{fig:FIG5_tem_cubic}. Clear evidence of each crystal phase is given by the fast Fourier transform (FFT) diffraction patterns of the observed atomic arrangements \cite{sun2004,bayram2014}. The wurtzite facets start growing at the edge of the SiC holes, where the graphene is present. GaN makes an overlaying contact with epitaxial graphene, enabling its use as a heat-dissipating back-electrode. The remaining fifth of the GaN nanomesas are smaller and flat-topped (see Figs.\ref{fig:FIG3_tem_all}b and \ref{fig:FIG4_tem_wurtzite}), and are mostly wurtzite. Evidence of their hexagonal crystal phase is given on Fig.\ref{fig:FIG4_tem_wurtzite}b. The very first layers of growth are cubic, as evidenced by their triangular faceting, which indicate a non-hexagonal symmetry: indeed they do not grow as six-faceted nanopyramids as in Refs.\cite{goh2009,puybaret2014,puybaret2015_2,puybaret2015_4}.
The very good uniformity of all the GaN nanomesas and the absence of threading dislocation emerging from the interface with SiC are all apparent. An immediate consequence is the absence of V-pits, which normally emerge from threading dislocations \cite{speck1996,wu1998}. 

The STEM analysis of the bigger GaN nanomesas, with large cubic central seed and out-growing wurtzite facets, in Fig.\ref{fig:FIG5_tem_cubic} clearly shows that cubic GaN grows along the [111] direction of the underlying 4H-SiC C-face. As a remark, right at the interface, the first layers of GaN continue the A-B-C-B-A stacking sequence of the [0001] axis of 4H-SiC, followed by GaN cubic growth along the [111] axis. Then, a transition region of a few atomic layer can be noticed in the high-angle annular dark field image (see Fig.\ref{fig:FIG5_tem_cubic}e): this intermediate zone, 2-3 nm from the edge of the SiC nano-hole, is a mix between both cubic and wurtzite phases, leading to a very thin and localized structure with stacking faults in an otherwise defect-free GaN nanomesa. Finally, at the edge of the SiC hole, right where graphene has not been etched away, GaN begins wurtzite growth (cf Figs.\ref{fig:FIG5_tem_cubic}d-e), with its own [0001] direction normal to the (\=111) plane of the underlying cubic GaN seed. This can be better understood by the following equivalence of lattice parameters: d\{111\}$_{cubic}$ = d\{0002\}$_{wurtzite}$. 

However, one fifth of the nanomesas seems to be fully hexagonal but are still triangular-shaped, similar to the larger nanomesas, instead of exhibiting the typical hexagon-base-pyramidal shape and its six (1\=101) or (1\=102) r-plane facets\cite{goh2009,puybaret2014,puybaret2015_2,puybaret2015_4}. An explanation is that the very first few atomic layers of GaN have a cubic structure, with a quick transition to growth of the wurtzite phase. The full STEM analysis in Fig.\ref{fig:FIG5_tem_cubic} also gives insight that can explain why every single GaN nanomesa on the SEM in Fig.\ref{fig:FIG2_graphene_nanosag_sem}b-c points up or down: the three (111) facets of the cubic phase grow along their three corresponding \{111\} axes, of which normal projections on the substrate plane are the three \{110\} directions. At the very beginning of growth, cubic GaN conforms to the underlying hexagonal SiC substrate, hence aligning its \{110\} axes with the underlying \{1100\} axes of the 4H-SiC. Therefore, the resulting triangular-shaped GaN nanomesas can only point up or down on Figs.\ref{fig:FIG2_graphene_nanosag_sem}b-c. 
The fact that graphene mask is extremely thin (5 to 8 atomic layers = 1.7 to 2.7 nm) compared to the typically used 100 nanometers of SiO$_2$ might also play a significant role at the beginning of growth. The polarity of the hexagonal SiC (Si-face or C-face), the composition of the mask, the geometry of the hole pattern are also factors to be considered for further study.

As a conclusion we have found that epitaxial graphene on silicon carbide can be used as a mask for NSAG of III-nitrides. This process is repeatable and is compatible with standard large-scale industrial electronics fabrication techniques. This is, to our knowledge, the first time graphene in any form (epitaxial on SiC, CVD, exfoliated) is used as a mask for NSAG. We show that a high-quality III-nitride material, with no threading dislocation, can be grown directly in contact of silicon carbide using large continuous films of patterned epitaxial graphene as a mask. The recovering of GaN over epitaxial graphene enables the use of the 2D carbon material as a high electron mobility and high thermal conductivity back-electrode. Moreover the technology presented here is fully developed on industry-grade and wafer-scalable SiC. 
The study also provides insight on the nucleation of the zinc-blende (non-polar) and wurtzite (polar) phases of GaN on 4H-SiC. At last the self-cleaning fabrication process presented here greatly simplifies the NSAG technique with the potential dual use of multilayer epitaxial graphene as growth mask and heat-dissipating back electrode for devices, in a cost-effective and solvent-free technology.


\section*{Materials and Methods}

\textit{Graphene growth}: Graphene was grown on the carbon-terminated face (000\=1) of research-grade semi-insulating CMP-polished 4H-SiC purchased from Cree Inc. by confinement-controlled sublimation of silicon carbide \cite{deheer2011_3}.
\\

\textit{Lithography}: Hydrogen silsesquioxane (HSQ) was used as the negative resist, and was patterned by a JEOL JBX-9300FS electron beam lithography system (4nm diameter Gaussian spot electron beam, 50kV-100kV accelerating voltage, 50pA-100nA current range). The resist was then developed in tetramethylammonium hydroxide (TMAH), and the exposed graphene was etched in an O$_2$ RIE plasma. HSQ was them removed by immersion in buffered oxide etch solution (BOE). Both TMAH and BOE are perfectly selective and harmless toward graphene and SiC.
\\

\textit{GaN growth}: GaN nanomesas were grown by metal organic vapor phase epitaxy (MOVPE). Trimethylgallium and ammonia were the gaseous precursors for gallium and nitrogen atoms, respectively. Growth was performed at temperature 1000$^\circ$C, under 80 Torr of N$_2$ as carrier gas.
\\

\textit{Characterizations}: Raman was performed by a  LabRam Horiba Jobin-Yvon system at 532 nm. Park System AFM was used for atomic force imaging. SEM was taken by a Zeiss Supra 55VP microscope. HRTEM and HAADF-STEM was performed with a Titan Thermis 200 at 200kV microscope with spherical aberration correction, HRTEM and HAADF detectors. An amorphous carbon layer was deposited on the samples to protect their surface for FIB preparation.

\section*{Acknowledgments}
The authors thank the French Lorraine Region, the National Science Foundation (DMR-0820382), the Air Force Office of Scientific Research for financial support, the Partner University Fund for a travel grant, the French National Research Agency (ANR), under the NOVAGAINS project (ANR-12-PRGE-0014- 02), and CNRS INCEPT PEPS project. C.B. also acknowledges partial funding from the Graphene Flagship EU grant N$^\circ$604391.

\bibliographystyle{ieeetr} 			
\bibliography{biblio}

\end{document}